\documentclass[preprint,aps]{revtex4}       
\usepackage{graphicx}
\begin{document}
\title{Rotating Neutron Stars in a Chiral SU(3) Model}
\author{S. Schramm}
\email{schramm@theory.phy.anl.gov}
\affiliation{Argonne National
Laboratory, 9700 S. Cass Avenue, Argonne IL 60439, USA}
\author{D. Zschiesche}
\affiliation{Institut f\"ur Theoretische Physik,
        Postfach 11 19 32, D-60054 Frankfurt am Main, Germany
}
\date{\today}

\begin{abstract}We study the properties of rotating neutron stars within a
generalized chiral SU(3)-flavor model. The influence of the
rotation on the inner structure and the hyperon matter content of the star
is discussed. We calculate the Kepler frequency and moments of
inertia of the neutron star sequences. An estimate for the braking
index of the associated pulsars is given.
\end{abstract}
\maketitle

 \section{Introduction}

 Neutron stars belong to the most important physical systems
 where strong interactions are ``tested'' at extreme dense
 conditions. In contrast to ultrarelativistic heavy-ion collisions
 where also high baryonic densities might be obtained by choosing the
 appropriate bombarding energy, neutron star
 matter exists at practically zero temperature - at least on a hadronic
 scale. Thus the understanding of neutron stars is the
 complementary task of what lattice gauge calculations can achieve for
 high-temperature, zero-density matter. In addition due to the fact that
 charge neutrality drives the stellar matter away from isospin-symmetric nuclear
 matter the study of neutron stars can give important clues for
 understanding the isospin-dependence of nuclear forces. A related regime
 of strong interactions can be experimentally
 investigated using secondary beams of nuclei going towards the neutron drip-line of
 nuclei.

 Compared to the harder task of determining masses or radii
 the most easily observable property of a neutron star is its
 rotational frequency through its pulsar radio signals emitted from the magnetic poles of the
 star\cite{prep}.
 Additional information can be obtained by the slow change of the pulsar frequency of an isolated
 neutron star
 with time signalling the slowing down of the rotation due to radiative loss of angular momentum.
 So far observational data in this respect are still scarce\cite{brak1,brak2,brak3,brak4}.
 \\ \indent
Various hadronic models have been used to describe the
structure of neutron stars. Going to central densities of a few times nuclear matter
saturation density  the inclusion of
hyperon degrees become essential \cite{glen}.
As it always will be difficult to get clear and unambiguous information
on the stellar interior from the relatively limited observational data
it is important to relate star properties with other observables like
neutron distribution radii in neutron-rich nuclei or multi-fragmentation
in heavy-ion collisions with different $N/Z$ nuclei.
It is therefore very useful to implement a model with a broader applicability including
 nuclear matter, finite nuclei and chiral symmetry restoration to be able to simultaneously
study different observables within a unified approach using the
same set of parameters for all calculations. In this paper we
follow  this direction by adopting a hadronic model incorporating
the underlying $SU_L(3)xSU_R(3)$ structure of  QCD, which we used
in  earlier studies of non-rotating neutron stars, nuclear matter
and finite nuclei/hypernuclei calculations
\cite{hanaus,ziesche,beckmann}.

Here we extend the static calculation to study the properties of rotating
neutron (hyper-) stars within the same approach. The outline of the article is as follows.
First we give a brief description of the ingredients of the hadronic model
that we implement in our calculations (a much more detailed discussion
can be found in \cite{paper3}).
After introducing the Tolman-Oppenheimer-Volkov (TOV) equations for the static star we
present
the extension to rotating stars following the treatment of Hartle and Thorne
\cite{hartle}.
In the third section we present some general features of the equation of state and its strange
matter contributions
and then look at the composition and gross properties of the rotating stars.
We conclude with an outlook of possible extensions of the current approach.

\section{The chiral model}

The chiral hadronic SU(3) lagrangian has the following basic structure
\begin{equation}
{\cal L} = {\cal L}_{\rm kin} + {\cal L}_{\rm BM} + {\cal L}_{\rm
BV} + {\cal L}_{\rm vec} + {\cal L}_0 + {\cal L}_{\rm SB} + {\cal
L}_{\rm lep} ~,
\end{equation}
consisting of interaction terms between baryons and spin-0 (BM) and spin-1 (BV) mesons
\begin{eqnarray}
{\cal L}_{\rm BM}+{\cal L}_{\rm BV} &=& -\sum_{i}
\overline{\psi}_i \left[ g_{i\sigma} \sigma + g_{i\zeta} \zeta + g_{i \omega}\gamma_0 \omega^0 +
g_{i \phi}\gamma_0 \phi^0
+ g_{N\rho} \gamma_0 \tau_3 \rho_0 \right] \psi_{i} ~, \nonumber \\
{\cal L}_{\rm vec} &=& \frac{1}{2}
m_{\omega}^2\frac{\chi^2}{\chi_0^2} \omega^2 + \frac{1}{2}
m_\phi^2\frac{\chi^2}{\chi_0^2} \phi^2 + \frac{1}{2}
\frac{\chi^2}{\chi_0^2} m_{\rho}^{2}\rho^2
+ g_4^4 (\omega^4 + 2 \phi^4 + 6 \omega^2 \rho^2 +\rho^4 )
\end{eqnarray}
summing over the baryonic octet, (p,n,$\Sigma^{-/0/+},\Xi^{-/0}$),
as well as interactions between the scalar mesons
\begin{eqnarray}
{\cal L}_0 &=& -\frac{1}{2} k_0 \chi^2 (\sigma^2+\zeta^2) + k_1
(\sigma^2+\zeta^2)^2 + k_2 ( \frac{ \sigma^4}{2} + \zeta^4)
+ k_3 \chi \sigma^2 \zeta \nonumber \\
& & - k_4 \chi^4 - \frac{1}{4}\chi^4 \ln \frac{ \chi^4 }{
\chi_0^4}
+\frac{\delta}{3}\ln \frac{\sigma^2\zeta}{\sigma_0^2 \zeta_0}~~.
\end{eqnarray}
An explicit symmetry breaking term mimics the QCD effect of non-zero current quark masses
\begin{eqnarray}
{\cal L}_{\rm SB} &=& -\left(\frac{\chi}{\chi_0}\right)^2
\left[m_\pi^2 f_\pi \sigma + (\sqrt{2}m_K^2 f_K -
\frac{1}{\sqrt{2}}
m_{\pi}^2 f_{\pi})\zeta \right]~~.
\end{eqnarray}
Finally, there are leptonic contributions from electrons and muons
\begin{eqnarray}
{\cal L}_{\rm lep} &=& \sum_{l=e, \mu} \overline{\psi}_l [i
\gamma_\mu\partial^\mu - m_l ]\psi_l ~.
\end{eqnarray}
${\cal L}_{\rm kin}$ in (1) contains the kinetic energy terms of
the hadrons. The general model incorporates the complete lowest
baryon and meson multiplets. In the previous formulae we only
considered the degrees of freedom relevant for a neutron star
calculation, the scalar field $\sigma$ and its $s\bar{s}$
counterpart $\zeta$ (which can be identified with the observed
$f_0$ particle), as well as the $\omega, \rho$ and $\phi$ vector
mesons. ${\cal L}_{\rm vec}$ generates the masses of the spin-1
mesons through the interactions with spin-0 mesons. The scalar
interactions ${\cal L}_0$ induce spontaneous chiral symmetry
breaking. Another scalar, isoscalar field incorporated in the
model, the dilaton $\chi$ simulates the breaking of the QCD scale
symmetry and can be identified with the gluon
condensate\cite{Schec}, for a longer discussion see
\cite{paper3}). The effective masses
$m^*(\sigma,\zeta)~=~g_{\sigma}~\sigma~+~g_{\zeta}~\zeta$ of the
baryons are generated through their coupling to the scalar fields,
which attain non-zero vacuum expectation values due to the
self-interactions ${\cal L_0}$ \cite{paper3}. As the field
strengths change in nuclear and neutron matter, the effective
masses of the baryon octet are shifted, too. The grand canonical
thermodynamic potential of the system at zero temperature can be
written as
\begin{equation}
\Omega/V = -{\cal L}_{\rm vec} - {\cal L}_0 - {\cal L}_{\rm SB}
-{\cal V}_{\rm vac} - \sum_i \frac{\gamma_i}{(2\pi)^3} \int d^3k
\left[E_i^*(k) - \mu_i^*\right] - \frac{1}{3} \sum_l
\frac{1}{\pi^2} \int \frac{ dk \: k^4}{\sqrt{k^2 + m_l^2}} .
\end{equation}
The chemical potentials for the baryons  read
\begin{equation}
\mu^*_i = \sqrt{k_{F,i}^2 + m_i^{* 2}} ~~,\quad  \mu_i = b_i\mu_n -
q_i\mu_e =  \mu^*_i + g_{i\omega}\omega_0 + g_{i\phi}\phi_0 +
g_{i\rho}\rho_0 ~,
\end{equation}
with $b_i$, $q_i$, and $k_{F,i}$ being the baryon number, charge,
and fermi momentum of the $i$th species. The energy density and
pressure are given by
\begin{eqnarray}
\varepsilon &=& \Omega/V + \sum_{k=i, l}\mu_k\rho_k \\
P &=& -\Omega/V.
\end{eqnarray}
By extremizing $\Omega/V$ one obtains self-consistent equations
for the meson fields in conjunction with demanding vanishing total
charge and $\beta$ equilibrium as the time scales involved allow
for weak interactions to take place.

\section{Stellar Equations}

In the case of the static neutron star we integrate the TOV equations\cite{Tol}:
\begin{eqnarray}
\label{tov1} \frac{dP(r)}{dr} &= - {(\epsilon + P) (4\pi r^3 P + m) \over r^2 (1-Y(r))}
\\
\label{tov2}
\frac{d m(r)}{dr} &= 4\pi r^2 \epsilon(r)~~,
\end{eqnarray}
where $P$ and $\epsilon$ are the pressure and energy density of
the nuclear matter. $m(r)$ is the gravitational mass of the star
inside of the radius $r$ and $Y(r) \equiv 2 m(r) / r$. Starting
with some initial values for $P$ and $\epsilon$, $P(r=0) \equiv
P_c$ and $\epsilon(r=0) = \epsilon_c(P_c)$, with $m(0) = 0$ one
can integrate equations (\ref{tov1}) and (\ref{tov2}) up to the
point of vanishing pressure, which defines the radius $R$ of the
star. Here the specific model of nuclear matter enters via the
equation of state $\epsilon(p)$. For the low-density tail of the
EOS we use a standard parametrization for the crust of the star as
discussed in \cite{hanaus}.
\\

Due to the rotation the star in general is deformed. Assuming axial symmetry
the metric of the rotating star can be written as:
\begin{equation}
\label{metrot}
ds^2 = \exp(2\nu) dt^2 - \exp(2\lambda) dr^2
- \exp(2 \mu)\left(d\theta^2 + \sin^2\theta (d\phi-\omega dt)^2\right)~~.
\end{equation}
The metric functions $\mu, \nu, \lambda$ depend on the radius $r$
and the polar angle $\theta$. If we set $\omega = 0$ we recover
the standard Schwarzschild metric for the static star with the
choice $\exp(2\mu)~=~r^2 ~~,~~ \exp(2\lambda)~=~1/(1-Y(r))$.
Because of the dragging of the local inertial frame $\omega(r)$
depends on the radius (but not to lowest order on the polar angle,
see \cite{hartle}) and is not equal to the rotational frequency
$\Omega$ of the star. Following \cite{hartle} and defining the
difference of the frequency of the local frame and $\Omega$ as
$\bar{\omega} \equiv \Omega - \omega$ demanding equilibrium
between pressure, centrifugal and rotational forces in the local
frame the differential equation for the metric functions
follows\cite{hartle}
\begin{equation}
\frac{d}{dr}\left[ j(r) \frac{d\bar{\omega}}{dr}+ \frac{4}{r}j(r)\bar{\omega}(r)\right]= 0
\end{equation}
with $j = \exp(-\nu -\lambda)$.
The rotation of the star generates a modification of the metric as shown in
(\ref{metrot}).
Using a multipole expansion of the metric functions, one obtains corrections that
up to quadrupole order read
\begin{eqnarray}
\exp(2\nu) &=& \exp(2\phi) \left( 1 + 2 \left(h_0 + h_2 P_2\left(\cos\theta\right)\right)\right)
\\
\exp(2\mu) &=& r^2 \left( 1 + 2 \left(v_2 - h_2 P_2\left(\cos\theta\right)\right)\right)
\\
\exp(2\lambda) &=& {1 \over 1 - Y(r)} \left( 1 + {2 \over r}
{m_0 + m_2 P_2\left(\cos\theta\right) \over 1 - Y(r)} \right)~~~,
\label{quadru}
\end{eqnarray}
where the static metric component $\phi$ is given by
\begin{equation}
{d \phi \over dr} = - (\epsilon +P)^{-1} {dP \over dr} \quad .
\end{equation}
The monopole coefficients $h_0$ and  $m_0$ obey
the differential equations ($r < R$):
\begin{eqnarray}
{d m_0 \over dr} &=& 4 \pi r^2 {\partial \epsilon \over \partial P} (\epsilon + P) p_0
+ {1 \over 12} j^2 r^4 \left({d\bar{\omega} \over dr}\right)^2
+ {8 \pi \over 3} r^4  j^2 {\epsilon + P \over 1 - Y} \bar{\omega}^2
\\
{d p_0 \over dr} &=& -{1 + 8\pi r^2 P \over r^2 ( 1-Y)^2} m_0
- 4\pi r {\epsilon + P \over 1 - Y} p_0 + {1\over 12} {r^3 j^2 \over 1 - Y}
\left({d\bar{\omega} \over dr}\right)^2 + {1 \over 3}
{d \over dr} \left({r^2 j^2 \bar{\omega}^2 \over 1 - Y}\right)
\\
h_0 &=& -p_0 + {r^2 \over 3} \bar{\omega}^2 \exp(-2\phi) + h_0^c~~~.
\end{eqnarray}
The value $h_0^c$ is fixed by requiring that $h_0(R) = - {m_0(R)
\over R (1-Y(R))}$ at the star's surface as can be inferred from
the large-r asymptotics for $h_0(r)$ \cite{weber}. Here $m_0$ and
$p_0$ are the monopole mass and pressure perturbation due to the
rotation of the star. The first terms where the actual deformation
of the star's shape show up are the quadrupole corrections $h_2,
m_2$ and $v_2$. They can be determined by
solving\cite{hartle,weber}
\begin{eqnarray}
{d v_2 \over dr} &=& - 2 {d \phi \over dr} h_2 + \left({1 \over r} + {d\Phi \over dr}\right)
\left\{ -{r^3 \over 3} {d j^2 \over dr} \bar{\omega}^2
+ \frac{j^2}{6} r^4 \left({d\bar{\omega} \over dr} \right)^2 \right\} \\
{d h_2 \over dr} &=& h_2 \left\{{d \phi \over dr} + {2 \over 1 - Y}
\left[ 2 \pi (\epsilon + P) - {m \over r^3} \right] / \left({d\phi \over dr} \right) \right\}
- {2 \over r^2 (1-Y)} / \left({d\phi \over dr} \right) v_2
\nonumber \\ &~+& {r^3 \over 6}\left(j {d\bar{\omega} \over dr} \right)^2
- \frac{1}{3} (r \bar{\omega})^2 {dj^2 \over dr}
\left\{ r{d\phi \over dr} + \left[ 2 r (1-Y) {d\phi \over dr} \right]^{-1} \right\}~~~~.
\end{eqnarray}
Integrating the equations and ensuring the correct asymptotics of vanishing corrections
at $r = 0$ and $r = \infty$ by adding the appropriate
solution of the homogeneous equations one obtains the quadrupole corrections to the metric.
The mass and pressure quadrupole corrections follow as \cite{hartle}
\begin{eqnarray}
m_2 &=&
r (1-Y) \left\{
{r^4 \over 6}\left(j {d\bar{\omega} \over dr} \right)^2
- \frac{r}{3} (r \bar{\omega})^2 {dj^2 \over dr} - h_2
\right\}
\\
p_2 &=&
- h_2 -{r^2 \over 3} \exp\left(-2\phi\right)\bar{\omega}^2~~~~.
\end{eqnarray}
After determining the metrical components we can then calculate radii, masses and moments of inertia
of the rotating solutions.

\section{Numerical Results}

We use the same parameter set (called ``C1'' in \cite{ziesche,paper3}) that was
also used in calculations of excited nuclear matter and finite nuclei.
\begin{figure}
\label{I3}
\includegraphics[width = 7cm]{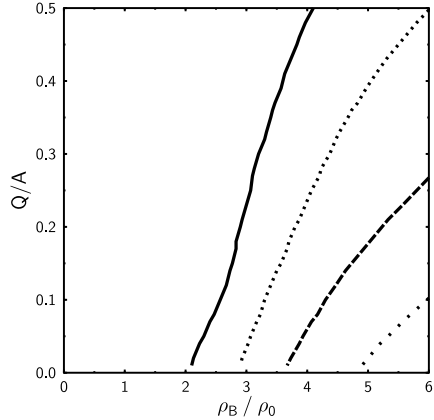}
\caption{Strangeness content of nuclear matter as function of
charge per baryon and density. The full line shows the onset of strangeness
in the system, the following lines mark the strangeness
fraction $f_s=0.1, 0.2$, and $0.3$ respectively, where $f_s$ is defined as number
of strange quarks per baryon}
\end{figure}
Figure 1 shows the result of a calculation of nuclear matter varying the
average charge of the system between symmetric nuclear matter ($Q/A = 0.5$) to
neutral hadronic matter ($Q = 0$). One can see how the onset of hypermatter is shifted
from neutral matter at a density of $\approx 2\rho_0$
to twice that value for symmetric nuclear matter.
Note that here no leptons are taken into account that shift the total baryonic charge to positive values
in the case of a neutron star. There, $\Lambda$ and $\Sigma$ states are populated
at densities higher than 2.6 $\rho_0$ \cite{hanaus}.
Looking again only at the baryonic contributions
\begin{figure}
\includegraphics[width = 7cm]{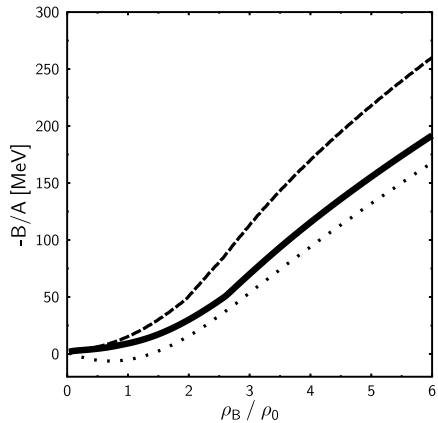}
 \label{eos} \caption{Binding energy of nuclear matter as function of
baryon density. The figure shows isospin symmetric matter (dashed line) and
neutral baryonic matter including strange baryons (solid line).
}
\end{figure}Figure~2 shows the
the energy per baryon $E/A - m_N$ as a function of density for total isospin per baryon
(3rd component) zero and -0.5. The isospin 0 curve exhibits the nuclear matter saturation properties.
The thick solid line shows the result of the baryonic energy inside the neutron star
(including leptons) as
calculated in the chiral model. One can see that for higher densities the equation of state resembles
more the isospin 0 nuclear matter results. In fact, in the extreme high-density limit
assuming similar population of
the baryonic octet the total isospin tends towards zero again.

\begin{figure}
\includegraphics[width = 7cm]{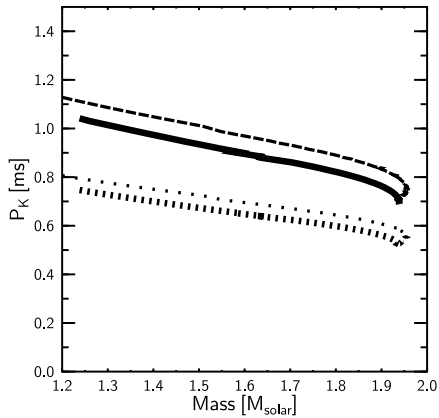}
 \label{kepler} \caption{Minimum rotational periods $P_K$ for stars with different central energy
density. The upper two curves are results for the chiral model (full line) and the TM1 relativistic mean
field model. The lower curves are the corresponding Newtonian estimates.}
\end{figure}
The maximum frequency $\Omega_K$ and the minimum rotational periods $P_s\equiv 2 \pi / \Omega_K$
of a rotating neutron star have been determined by
integrating the self-consistent equation \cite{weber}
\begin{equation}
\bar{\omega} = V(\Omega) \exp\left[\nu(\Omega) - \mu(\Omega)\right]
\end{equation}
solving for $\Omega$ at the equator of the star, where the mass
shedding occurs first, for different central energy densities and
total masses of the star. The resulting values are shown in
Figure~3. Compared to the Newtonian value of the Kepler period
(dashed lines) there is an increase in $P_K$ of about 25\% due to
the general relativity corrections. This result is in accordance
with other calculations of different neutron star models, which
show similar results \cite{weber2}. In the plot also results for a
relativistic mean-field parametrization (TM1 \cite{tm1}) are
shown, which was extended to include hyperons (see \cite{hanaus}).
The difference of the Kepler frequency for both parameterizations
is small. Note that the maximum gravitational mass of the neutron
star that was about 1.64 M$_{\odot}$ in case of the static
solution \cite{hanaus} is now increased to 1.94 solar masses due
to the additional rotational contribution. Looking at the
frequency dependence by spinning up the neutron star one gets the
increasing neutron star mass as shown in Figure~4. One TM1 result
for a star with M($\Omega$=0)~=~1.48M$_{\odot}$ is shown for
comparison. In this case one can see a distinctly sharper rise of
the mass of the star with rotational frequency.
\begin{figure}
\includegraphics[width = 7cm]{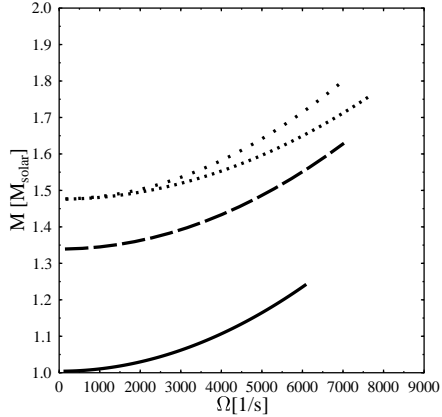}
 \label{momega} \caption{Mass of different neutron star as function of rotational frequency.
The sharply rising dotted curve show a result for the TM1 parameterization.
}
\end{figure}
For the chiral case one can observe the increase of the mass by up
to 30\% at the mass-shedding frequency.

Similarly looking at the monopole corrections one can calculate the increase of
the radius of different star solutions due to rotation averaged over the star's surface.
\begin{figure}
\includegraphics[width = 7cm]{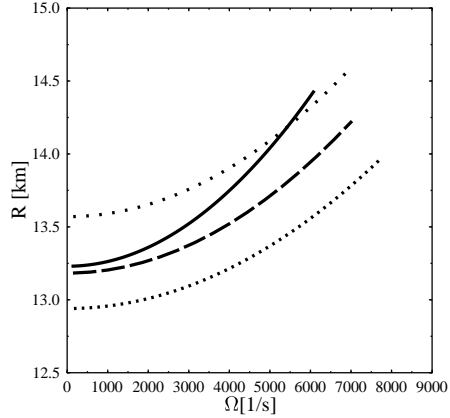}
 \label{romega} \caption{Radius of different neutron star (cf. Fig.~4) as function of rotational frequency.
}
\end{figure}
Figure~5 shows
that there is an increase of the radius by up to 10\% close to $\Omega_K$. Again, the TM1
model shows a more drastic change with rotation.
From calculating the quadrupole coefficients Eqs.~(\ref{quadru}) one can determine the
excentricity $\epsilon_{ex}$ of the star defined via the ratio of the surface radius at the equator
over the pole radius:
\begin{equation}
\epsilon_{ex} = \sqrt{1 - R_{pole}^2 / R_{equator}^2}
\end{equation}
\begin{figure}
\includegraphics[width = 7cm]{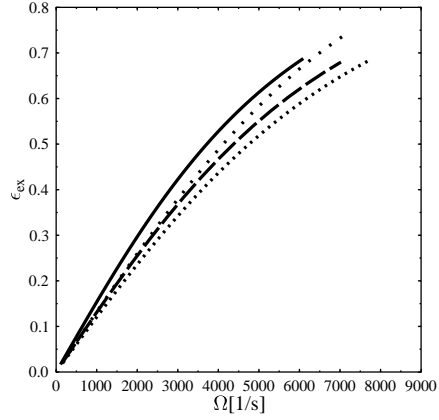}
 \label{exomega} \caption{Excentricity of the deformed rotating neutron star as function of rotational frequency.
}
\end{figure}

Figure~6 shows that there is a substantial deformation of the star. This has also
an impact on the internal structure of the star depending on the polar angle. This
is shown in Figure~7.
\begin{figure}
\includegraphics[width = 7cm]{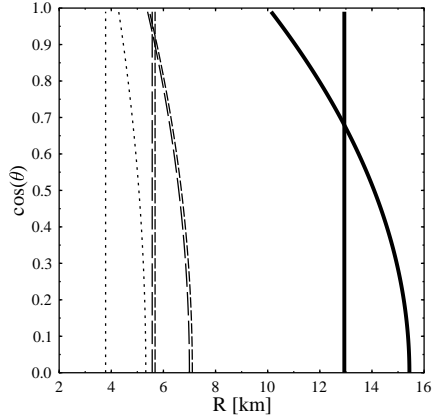}
 \label{structure} \caption{Inner structure of the neutron star. The onset
 of hyperon population in the interior of the neutron star as function of depth
 and polar angle is
 shown for $\Lambda$ (long-dashed lines), $\Sigma^-$ (dashed lines) and
 $\Xi^-$ particles (dotted lines) is
 shown. The full line represents the star's surface.
 For comparison the thresholds and the radius for the corresponding static star
 are shown (straight lines).
}
\end{figure}
Here the onset of the various hyperons contained in the star
$\Sigma^-, \Lambda$ and $\Xi-$ are shown for a typical static star with a Mass of 1.475 M$_\odot$
(vertical lines) compared to the same solution corrected for rotational effects at its Kepler frequency.
\\
Integrating the inertia of the star by using the general expression \cite{weber2}
\begin{equation}
I = 2\pi \int d\theta dr \exp\left\{\lambda+\mu+\nu+\psi\right\} (\epsilon + P)
\left(\exp\left[2\nu-2\psi\right] - \bar{\omega}^2\right) {\bar{\omega} \over \Omega}
\end{equation}
and again taking into account the corrections to the metric up to order $\Omega^2$
one gets the moment of inertia of the star as shown in Fig.~8. In this case
the mass of the non-rotating star is 1.48 solar masses.
\begin{figure}
\includegraphics[width = 7cm]{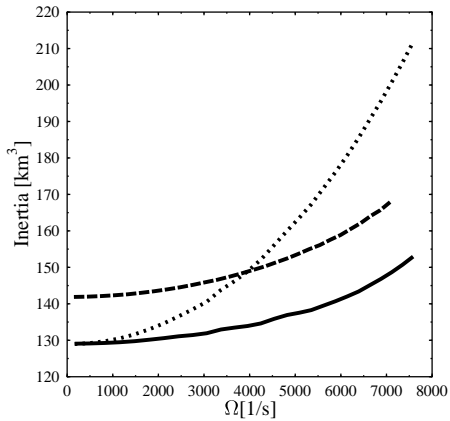}
 \label{inertia} \caption{Moment of inertia of the rotating star as function
 of frequency. A fixed value of $M_B$ is assumed. The result of the chiral model (solid line),
 the TM1 calculation (dashed lines) and in comparison the inertia of the star neglecting
 the deformation (dotted line) in the case of the chiral model are shown.}
\end{figure}
In this figure
we have kept the baryonic number of the star constant at $M_B = 1.22\cdot 10^{57} \sim (M_B)_{solar}$
and changed the rotational
frequency. The curve therefore shows the change in the moment of inertia during the spin-down of
a pulsar, the energy-loss of the star generated by electromagnetic and gravitational radiation.
Assuming a simple power-law of the energy loss $\Delta E \sim - E^n_o$ where $n_o=3$ and $n_o=5$
in the case
of magnetic and gravitational radiation, respectively. With this assumption one can determine
the frequency-dependent braking index $n(\Omega) = n_o - \Delta n(\Omega)$ with
\begin{equation}
\Delta n =\frac{3 I'\Omega + I''\Omega^2}{2I+I'\Omega}
\end{equation}
where the prime denotes the derivative with respect to frequency.
Figure~9 shows the shift of the braking index.
\begin{figure}
\includegraphics[width = 7cm]{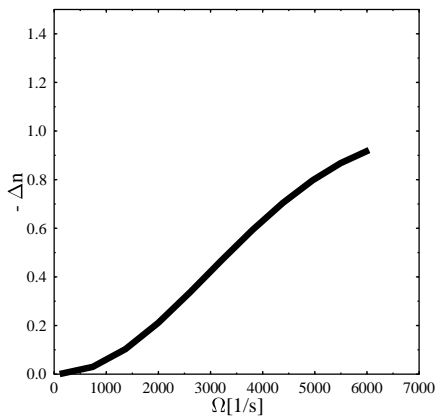}
 \label{braking} \caption{Deviation of the braking index from an assumed power law.
 In the case of magnetic radiation the index is reduced by a factor of up to 1/3.}
\end{figure}
There is a substantial reduction in the case of fast rotating
stars that has to be taken into account for an analysis of the
time evolution of the star. However, one cannot see significant
structures as for instance a back-bending shape as seen in moments
of inertia of atomic nuclei and also suggested for neutron
stars\cite{backbending}. This is not very surprising, since the
chiral model does not show a phase transition for low temperatures
and at high densities (see, however, the discussion in the
concluding section). Note that in this case at maximum rotation
the star starts  with a central density of 2.3 times nuclear
matter density, which is below the threshold of the occurrence of
hyperons. Below a frequency $\Omega \sim 5600/$s $(P \sim 1.1 $ms)
hypermatter becomes energetically more favorable.

\section{Conclusion}

We have used a chiral SU(3) model for the calculation of  rotating
neutron stars. The model parameters have been fitted to reproduce
hadronic masses and saturation properties of groundstate nuclear
matter. Within the same model and with the same parameters also
nuclei and hypernuclei have been described quite successfully. The
resulting star shapes show a significant excentricity which can
also be seen from its hypermatter distribution in the interior of
the star. The moment of inertia shows a rather strong dependence
on the rotational frequency, which has to be taken into account
for analyzing the spinning-down and related age-estimates of
pulsars. No back-bending of the inertia occurs due to the lack of
a phase transition of the chiral model for high densities and low
temperatures. In general the comparison with a quite different
model that contains hyperons shows the robustness of the results
for the rotating star as well as the relative insensitivity of the
star properties on very different models with quite distinct
hyperon content\cite{hanaus}. In this context a discussion of the
possible effects of populating higher baryonic resonances in the
neutron star should be performed. Those additional degrees of
freedom can change the high-density behavior of the system
significantly, generating a different structure of the
time-evolution of the rotating star. Such a calculation might rule
in or out specific theoretical descriptions of hadronic resonances
by studying their effect on neutron star properties. Work in this
direction is in progress~\cite{future}.

\begin{acknowledgments}
This work
was supported in part by the U.S. Department of Energy, Nuclear Physics
Division (Contract No. W-31-109-Eng-38). The authors thank M. Hanauske
and S. Pal for valuable discussions.
\end{acknowledgments}

\end{document}